\documentclass[twocolumn,english,pra,showpacs]{revtex4-1}
\usepackage[T1]{fontenc}
\usepackage[latin9]{inputenc}
\usepackage{color}
\usepackage{babel}
\usepackage{amsmath}
\usepackage{amssymb}
\usepackage{esint}
\usepackage{xargs}[2008/03/08]
\usepackage[unicode=true,pdfusetitle,
 bookmarks=true,bookmarksnumbered=false,bookmarksopen=false,
 breaklinks=false,pdfborder={0 0 0},backref=false,colorlinks=true]
 {hyperref}

\makeatletter

\usepackage{xargs}[2008/03/08]

\usepackage{babel}

\makeatother

\begin{document}

\title{Quantum metrology for a general Hamiltonian parameter}

\author{Shengshi Pang}

\email{shengshp@usc.edu}

\author{Todd A. Brun}

\email{tbrun@usc.edu}

\affiliation{Department of Electrical Engineering, University of Southern California,
Los Angeles, California 90089, USA}
\begin{abstract}
Quantum metrology enhances the sensitivity of parameter estimation
using the distinctive resources of quantum mechanics such as entanglement.
It has been shown that the precision of estimating an overall multiplicative
factor of a Hamiltonian can be increased to exceed the classical limit,
yet little is known about estimating a general Hamiltonian parameter.
In this paper, we study this problem in detail. We find that the scaling
of the estimation precision with the number of systems can always
be optimized to the Heisenberg limit, while the time scaling can be
quite different from that of estimating an overall multiplicative
factor. We derive the generator of local parameter translation on
the unitary evolution operator of the Hamiltonian, and use it to evaluate
the estimation precision of the parameter and establish a general
upper bound on the quantum Fisher information. The results indicate
that the quantum Fisher information generally can be divided into
two parts: one is quadratic in time, while the other oscillates with
time. When the eigenvalues of the Hamiltonian do not depend on the
parameter, the quadratic term vanishes, and the quantum Fisher information
will be bounded in this case. To illustrate the results, we give an
example of estimating a parameter of a magnetic field by measuring
a spin-$\frac{1}{2}$ particle and compare the results for estimating
the amplitude and the direction of the magnetic field.

\global\long\def\ng{\overrightarrow{n(g)}}
 \global\long\def\ngp{\overrightarrow{n^{\prime}(g)}}
 \global\long\def\ngo{\overrightarrow{n_{0}^{\prime}(g)}}
 \global\long\def\mg{\overrightarrow{m(g)}}
 \global\long\def\lg{\overrightarrow{l(g)}}
 \global\long\def\i{\mathrm{i}}
 \global\long\def\d{\mathrm{d}}
 \global\long\def\eg{\overrightarrow{\eta(g)}}
 \global\long\def\oe{\overrightarrow{X}}
 \global\long\def\oy{\overrightarrow{Y}}
 \global\long\def\oh{\overrightarrow{H}}
 \global\long\def\tr{\mathrm{Tr}}
 \global\long\def\bo{\boldsymbol{0}}
 \global\long\def\e{\mathrm{e}}
 \global\long\def\t{\mathrm{total}}
 \global\long\def\est{\mathrm{est}}
 \global\long\def\psi{|\Psi\rangle}
 \global\long\def\psc{\langle\Psi|}
 \newcommandx\qf[1][usedefault, addprefix=\global, 1=]{F_{g#1}^{(Q)}}
 \newcommandx\hg[1][usedefault, addprefix=\global, 1=]{H_{g#1}}
 \global\long\def\psf#1{|\Psi_{f}(#1)\rangle}
 \global\long\def\qft{F_{g,\t}^{(Q)}}
 \global\long\def\sh{\mathcal{H}_{g}}
 \newcommandx\h[2][usedefault, addprefix=\global, 1=, 2=]{h_{g#2}^{#1}}
 \global\long\def\n{n_{g}}

\end{abstract}


\maketitle

\section{Introduction}

Quantum metrology \cite{review 1,review 2} is a scheme that uses
entanglement to increase the precision of parameter estimation by
quantum measurements beyond the limit of its classical counterpart.
In classical parameter estimation, the estimation precision scales
as $\nu^{-\frac{1}{2}}$, where $\nu$ is the number of rounds of
measurement. The scaling can be rewritten as $(N\nu)^{-\frac{1}{2}}$,
where $N$ is the number of qubits used in each round, for parameter
estimation by quantum measurements if the $N$ qubits are not entangled.
This scaling is often termed as the \emph{standard quantum limit}
(SQL) \cite{SQL}, which characterizes the precision limit of quantum
measurements in the presence of the shot noise. A more fundamental
imprecision of quantum measurement originates from the Heisenberg
uncertainty principle, which is one of the most fundamental properties
of quantum mechanics, due to the probabilistic nature of quantum measurements.
Research in quantum metrology has shown that with the assistance of
$n$-qubit entanglement, the optimal scaling of the estimation precision
can be raised to $N^{-1}\nu^{-\frac{1}{2}}$, i.e., the \emph{Heisenberg
limit}, implying an improvement of $N^{\frac{1}{2}}$ over the SQL.

Quantum metrology is rooted in the theory of quantum estimation, which
was pioneered by Helstrom \cite{Helstrom} and Holevo \cite{Holevo}
who proposed the parameter-based uncertainty relation. Braunstein
\emph{et al.} \cite{quantum fisher inf 1,quantum fisher inf 2} developed
that theory from the view of the Cramér-Rao bound \cite{Cramer-Rao bouond},
which characterizes how well a parameter can be estimated from a probability
distribution, and obtained the optimal Fisher information over different
quantum measurement schemes for a given parameter-dependent quantum
state. This is often called \emph{quantum Fisher information.}

Given the importance of precision measurement in different fields
of physics and engineering, the quantum Fisher information has attracted
great interest from researchers. Giovannetti \emph{et al.} \cite{review 1}
found that the scaling of the quantum Fisher information has an $N^{\frac{1}{2}}$
improvement compared to its classical counterpart if an $N$-qubit
maximally entangled state is used. This stimulated the emergence of
quantum metrology, which has been applied to different quantum systems
to raise the precision of measurements.

The optimality of quantum metrology in terms of the scaling of the
measurement precision was proved in \cite{optimality 1} for different
initial states and measurement schemes, and also by \cite{optimality 2}
from the viewpoint of the query complexity of a quantum network. Moreover,
when there is interaction among the $N$ entangled qubits, the measurement
precision can be further increased to beyond the Heisenberg limit
\cite{interaction 1,interaction 2,interaction 3,interaction 4,interaction 5,interaction 6}.

Many applications of quantum metrology have been found, including
quantum frequency standards \cite{frequency measurement,frequency measurement 2},
optical phase estimation \cite{optical phase 0,optical phase 1,optical phase 2,optical phase 3,optical phase 4,optical phase 5,optical phase 6},
atomic clocks \cite{quantum clock 1,quantum clock 2,quantum clock 3,quantum clock 4,quantum clock 5},
atomic interferometers \cite{frequency measurement 2,atomic interferometer,atomic interferometer-1},
quantum imaging \cite{atomic interferometer-1,quantum imaging}, and
quantum-enhanced positioning and clock synchronization \cite{position measurement}.
The quantum Fisher information has also been studied in open systems
\cite{open system 0,open system 1,open system 2,open system 3,open system 4,open system 5,open system 6},
along with growing research on protocols assisted by error correction
\cite{qec 1,qec 2,qec 3}. Moreover, quantum metrology with nonlinear
Hamiltonians has received considerable attention \cite{interaction 4,nonlinear 1,nonlinear 2,nonlinear 3,nonlinear 4,nonlinear 5,nonlinear 6,nonlinear 7,nonlinear 8,nonlinear 9}.
For reviews of the field of quantum metrology, refer to \cite{review 1,review 2}.

Studies of quantum metrology have mainly focused on the precision
of measuring an overall multiplicative factor of a Hamiltonian, e.g.,
the parameter $g$ in a Hamiltonian $gH$, a setting particularly
suitable for enhancing phase or frequency estimation in devices such
as optical interferometers or atomic spectroscopes. However, generally
speaking, a parameter can appear in a more general form in a Hamiltonian,
not necessarily as an overall multiplicative factor. For example,
the parameter can appear with different orders in the eigenvalues
of the Hamiltonian, or even in the eigenstates of the Hamiltonian.
An understanding of the quantum limits in estimating this kind of
general parameter is emerging (e.g., \cite{Brody} from the view of
information geometry), but is still rather limited so far, which restricts
the potential range of applications of quantum mechanics to metrology.

This paper extends quantum metrology to estimating a general parameter
of a Hamiltonian. We will show that the optimal scaling of the measurement
precision with the number $N$ of systems is still $N^{-1}$, but
the time scaling will be different. In detail, it will be shown that
the quantum Fisher information can generally be divided into two parts:
one is linear in the time $t$, corresponding to the variation of
the eigenvalues, and the other is oscillatory, corresponding to the
variation of the eigenvectors of the Hamiltonian. The oscillating
part is bounded no matter how long the time $t$ is. We will obtain
an upper bound on the Fisher information for the general case.

The study of this problem will extend the current knowledge of quantum
metrology to a more general case, and more kinds of precision measurements
will benefit from this extension, especially those that go beyond
phase or frequency measurement. For instance, as we show as an example
in this paper, it can enhance the precision of measuring the direction
of a magnetic field by a spin-$\frac{1}{2}$ system, which is useful
for calibrating the field. Therefore, the results of this paper will
be useful to both theory and experiments in quantum metrology.

\section{Preliminary}

Let us first review some concepts of the estimation theory and their
quantum counterparts. The task of parameter estimation is to determine
a parameter from a set of data which depends on the parameter. A general
procedure for estimating a parameter is as follows: first acquire
a set of data $x_{1},\cdots,x_{\nu}$ which obey a probability distribution
dependent on the parameter $f_{g}(x)$, where $g$ is the parameter
to estimate; then estimate $g$ from $x_{1},\cdots,x_{\nu}$ by a
certain estimator, and obtain the estimated value $g_{\est}(x_{1},\cdots,x_{\nu})$.
While there are many different estimation strategies, such as the
method of moments and maximum-likelihood estimation, the performances
of those strategies differ. One of the most important benchmarks of
a strategy is the estimation precision, which is usually characterized
by the estimation error \cite{quantum fisher inf 1}: 
\begin{equation}
\delta g\equiv\frac{g_{\est}}{|\d\langle g_{\est}\rangle_{g}/\d g|}-g,
\end{equation}
where the factor $|\d\langle g_{\est}\rangle_{g}/\d g|$ is to eliminate
the local difference in the units between the estimator and the real
parameter for different $g$. If the estimation procedure is repeated
many times, the estimated value $g_{\est}$ may have fluctuations.
So an appropriate measure to quantify the performance of an estimator
is the root-mean-square error of the estimation results $\langle(\delta g)^{2}\rangle^{\frac{1}{2}}$.
A cornerstone of the classical theory of parameter estimation is the
Cramér-Rao bound \cite{Cramer-Rao bouond}, which bounds the precision
limit of an estimator by the following relation: 
\begin{equation}
\langle(\delta g)^{2}\rangle\geq\frac{1}{\nu F(g)}+\langle\delta g\rangle^{2},\label{eq:3}
\end{equation}
where $F_{g}$ is the Fisher information defined as 
\begin{equation}
F_{g}=\int[\partial_{g}\ln f_{g}(x)]^{2}f_{g}(x)\d x.
\end{equation}
The second term on the right side of \eqref{eq:3}, $\langle\delta g\rangle^{2}$,
characterizes the bias of the estimator. If the estimator is unbiased,
i.e., $\langle g_{\est}\rangle_{g}=g$, then $\langle\delta g\rangle=0$.

The achievability (or the tightness) of the Cramér-Rao bound \eqref{eq:3}
is addressed by the Fisher theorem. Fisher proved that for asymptotically
large $\nu$, the Cramér-Rao bound can always be achieved by maximum-likelihood
estimation (MLE) and the estimation result is unbiased. Because of
this property, MLE has been widely adopted in parameter estimation
protocols.

In quantum metrology, one measures a parameter-dependent state, say
$\rho_{g}$, to estimate $g$. The process of a quantum metrology
protocol splits into two stages. First, measure the state in some
basis {[}or, more generally, perform a positive operator-valued measure
(POVM) on it{]} and record the measurement result. When such a measurement
is repeated $\nu$ times for the same $\rho_{g}$, we will acquire
$\nu$ measurement results. These results depend on $g$, so they
can be used as sample data to estimate $g$. The second stage is estimating
the parameter $g$ based on the measurement results by some appropriate
estimation strategy. The precision of the estimation is bounded by
\eqref{eq:3} as usual. The complexity of quantum metrology comes
from the many different choices of the measurements (or POVMs). Different
choices lead to different precisions of the estimation results. The
aim of quantum metrology is to increase the estimation precision by
optimizing the measurement basis (or POVM).

Braunstein and Caves obtained the optimal Fisher information over
all POVMs for a given $\rho_{g}$ \cite{quantum fisher inf 1,quantum fisher inf 2},
which is called the quantum Fisher information, through the logarithmic
derivative $L_{g}$ \cite{Holevo}: 
\begin{equation}
\qf=\tr(L_{g}^{\dagger}\rho_{g}L_{g}).\label{eq:4}
\end{equation}
The logarithmic derivative $L_{g}$ has several different but equivalent
definitions. The most common is the symmetric logarithmic derivative
(SLD), defined as $\partial_{g}\rho_{g}=(L_{g}\rho_{g}+\rho_{g}L_{g})/2$.
$L_{g}$ in this definition is Hermitian, and the quantum Fisher information
$\qf$ \eqref{eq:4} can be simplified to $\tr(\rho_{g}L_{g}^{2})$.
In the eigenbasis of $\rho_{g}$, an explicit form of $L_{g}$ can
be found: 
\begin{equation}
L_{g}=2\sum_{i,j}\frac{\langle\eta_{i}|\partial_{g}\rho_{g}|\eta_{j}\rangle}{\eta_{i}+\eta_{j}}|\eta_{i}\rangle\langle\eta_{j}|,
\end{equation}
where the $\eta_{i}$'s are the eigenvalues of $\rho_{g}$ and the
$|\eta_{i}\rangle$'s are the corresponding eigenstates.

In the current literature of quantum metrology, most research interest
has been focused on estimating an overall multiplicative factor of
a Hamiltonian, for example, estimating $g$ in a Hamiltonian $gH$.
Usually an initial pure state $\psi$ is used to undergo evolution
by the Hamiltonian so that 
\begin{equation}
\rho_{g}=\exp(-\i gtH)\psi\psc\exp(\i gtH).
\end{equation}
In such a case, the quantum Fisher information $\qf$ can be simplified
to 
\begin{equation}
\qf=4t^{2}\psc\Delta H^{2}\psi.
\end{equation}

It can be proved \cite{optimality 1} that $\psc\Delta H^{2}\psi_{\max}=\frac{1}{4}(E_{\max}-E_{\min})^{2}$,
where $E_{\max}$ and $E_{\min}$ are the maximal and minimal eigenvalues
of $H$, respectively. Since $E_{\max}$ and $E_{\min}$ grow linearly
with the number of systems $N$, $\qf\propto N^{2}$, which is the
origin of the $\sqrt{N}$ improvement of the precision scaling in
quantum metrology compared with the SQL.

\section{$N$ scaling of quantum Fisher information}

Now we turn to the major problem of this paper. We are interested
in quantum metrology for a general parameter in a Hamiltonian. Both
the eigenvalues and the eigenstates of the Hamiltonian may depend
on the parameter. We mainly consider the scaling of the quantum Fisher
information with the number of the systems $N$ in this section, and
leave the more general results for the next section.

We first introduce the general framework of how to derive the quantum
Fisher information of estimating a Hamiltonian parameter. Suppose
the Hamiltonian of a single system is $\hg$, the initial state of
the system is $\rho_{0}$, and the parameter we want to estimate is
$g$. After the evolution under the Hamiltonian, the state of the
system becomes $\rho_{g}=U_{g}\rho_{0}U_{g}^{\dagger}$, where $U_{g}=\exp(-\i t\hg).$
The sensitivity of $\rho_{g}$ to the parameter $g$ can be characterized
by the generator of the local parameter translation from $\rho_{g}$
to $\rho_{g+\d g}$, where $\d g$ is an infinitesimal change of $g$.

In detail, when $g$ is changed to $g+\d g$, $\rho_{g}$ is updated
to $\rho_{g+\d g}=U_{g+\d g}\rho_{0}U_{g+\d g}^{\dagger}$. Since
$U_{g+\d g}\approx U_{g}+\partial_{g}U_{g}\d g$, the translation
from $\rho_{g}$ to $\rho_{g+\d g}$ can be written as 
\begin{equation}
\begin{aligned}\rho_{g+\d g} & \approx(U_{g}+\d g\partial_{g}U_{g})\rho_{0}(U_{g}^{\dagger}+\d g\partial_{g}U_{g}^{\dagger})\\
 & =(I+\d g(\partial_{g}U_{g})U_{g}^{\dagger})U_{g}\rho_{0}U_{g}^{\dagger}(I+\d gU_{g}\partial_{g}U_{g}^{\dagger})\\
 & \approx\exp(-\i\h\d g)\rho_{g}\exp(\i\h\d g),
\end{aligned}
\end{equation}
where 
\begin{equation}
\h=\i(\partial_{g}U_{g})U_{g}^{\dagger}.\label{eq:15}
\end{equation}
So, $\h$ is the generator of parameter translation with respect to
$g$, and the subscript $g$ is to indicate that this generator is
\emph{local} in $g$. It can be shown \cite{quantum fisher inf 1,quantum fisher inf 2}
that when the initial state of the system is a pure state $\psi$,
the quantum Fisher information of the evolved state $|\Psi_{g}\rangle=U_{g}\psi$
for the parameter $g$ is 
\begin{equation}
\qf=4\langle\Psi_{g}|\Delta\h[2]|\Psi_{g}\rangle,\label{eq:13}
\end{equation}
And the variance of $\h$ is maximized when $|\Psi_{g}\rangle=\frac{1}{\sqrt{2}}(|\lambda_{\max}(\h)+\e^{\i\varphi}|\lambda_{\min}(\h)\rangle)$
($\e^{\i\varphi}$ is an arbitrary phase) \cite{optimality 1}, so
the maximal quantum information is 
\begin{equation}
\qf[,\max]=(\lambda_{\max}(\h)-\lambda_{\min}(\h))^{2},\label{eq:6}
\end{equation}
where $\lambda_{\max}(\h)$ and $\lambda_{\min}(\h)$ are the maximal
and minimal eigenvalues of $h$, respectively.

When there are $N$ systems, the total Hamiltonian is $\hg[,{\rm total}]=\hg[,1]+\cdots+\hg[,N]$,
where $\hg[,i]$ is the Hamiltonian for the $i$th system alone, i.e.,
$\hg[,i]=I^{\otimes i-1}\otimes\hg\otimes I^{\otimes N-i}$. Since
$[H_{i},H_{j}]=0,\,\forall i,j=1,\cdots,N$, we have 
\begin{equation}
\begin{aligned}h_{g,{\rm total}} & =\i\frac{\partial\e^{-\i t\hg[,{\rm total}]}}{\partial g}\e^{\i t\hg[,{\rm total}]}\\
 & =\h[][,1]+\cdots+\h[][,N].
\end{aligned}
\end{equation}
As $\hg[,1],\cdots,\hg[,N]$ are the same Hamiltonian on different
systems, it is obvious that 
\begin{equation}
\begin{aligned}\lambda_{\max}(\h[][,{\rm total}]) & =N\lambda_{\max}(\h),\\
\lambda_{\min}(\h[][,{\rm total}]) & =N\lambda_{\min}(\h).
\end{aligned}
\end{equation}
So according to \eqref{eq:6}, 
\begin{equation}
\max\qf[,{\rm total}]=N^{2}\qf[,\max],\label{eq:14}
\end{equation}
where $\qf[,{\rm total}]$ is the total quantum Fisher information
of the $N$ systems.

Equation \eqref{eq:14} is interesting since it implies that the optimal
scaling of the total Fisher information using $N$ systems can always
reach $N^{2}$, which beats the classical scaling limit and is universal
for estimating an arbitrary parameter in the Hamiltonian. Of course,
if there are interactions among the $N$ systems, the optimal scaling
of the Fisher information may be even higher, which has been found
for estimating an overall multiplicative factor of a Hamiltonian \cite{interaction 1,interaction 3}.
In that case, the total Hamiltonian becomes $H_{g,\t}=\sum_{i_{1},\cdots,i_{k}}H_{g,\langle i_{1},\cdots,i_{k}\rangle}$
if there are $k$-body interactions among the $N$ systems. Obviously,
the total Hamiltonian can grow nonlinearly with $N$ in general, so
the quantum Fisher information may increase faster than $N^{2}$.
Such a case is beyond the scope of this paper and we do not consider
it in detail here.

\section{Quantum Fisher information for general Hamiltonian parameters}

In this section, we study quantum metrology for general Hamiltonian
parameters in detail. We consider only single systems here, and focus
on the time scaling of the quantum Fisher information, since the scaling
with the number of systems was treated in the previous section. It
can be seen from Eq. \eqref{eq:6} that the key to the quantum Fisher
information $\qf$ is the generator $\h$ \eqref{eq:15} of the local
parameter translation from $U_{g}$ to $U_{g+\d g}$, so our main
effort is to derive $\h$ in the following.

\subsection{Result for $t\ll1$}

First, we study the derivative of $\exp(-\i tH(g))$ with respect
to $g$ which is needed in \eqref{eq:15}. This derivative is nontrivial,
since $\hg$ does not commute with $ $$\partial_{g}\hg$ in general.
To obtain this derivative, we start from an integral formula for the
derivative of an operator exponential \cite{operator derivative}:
\begin{equation}
\begin{aligned} & \frac{\partial\exp[-\i\beta H(\lambda)]}{\partial\lambda}\\
= & -\i\int_{0}^{\beta}\exp[-\i\mu H(\lambda)]\frac{\partial H(\lambda)}{\partial\lambda}\exp[(\i\mu-\i\beta)H(\lambda)]\d\mu,
\end{aligned}
\end{equation}
where $\mu,\beta\in\mathbb{R}$. By this formula and according to
the definition of $\h$ \eqref{eq:15}, we get 
\begin{equation}
\h=\int_{0}^{t}\exp(-\i\mu\hg)\partial_{g}\hg\exp(\i\mu\hg)\d\mu.\label{eq:1-1}
\end{equation}

When $t\ll1$, the first-order approximation of \eqref{eq:1-1} is
\[
\h\approx t\partial_{g}\hg.
\]
As shown in Appendix B, we can get an upper bound for the quantum
Fisher information in this case, 
\begin{equation}
\qf[,\max]\leq\frac{t^{2}}{2}\tr(\partial_{g}\hg)^{2}.
\end{equation}

\subsection{Result for general $t$}

For larger $t$, direct calculation of the integral in \eqref{eq:1-1}
is not easy. Of course, one can use the Baker-Campbell-Hausdorff formula
to expand the integrand, but that will yield an infinite series that
is difficult to treat. So we resort to a different approach to work
out $\h$, which was first proposed in \cite{operator derivative}.

Denote the integrand of \eqref{eq:1-1} as $Y(\mu)$: 
\begin{equation}
Y(\mu)=\exp(-\i\mu\hg)\frac{\partial\hg}{\partial g}\exp(\i\mu\hg).\label{eq:8}
\end{equation}
The derivative of $Y(\mu)$ with respect to $\mu$ satisfies 
\begin{equation}
\frac{\partial Y}{\partial\mu}=-\i[\hg,Y],\label{2}
\end{equation}
and the initial condition is $Y(0)=\partial_{g}\hg$.

To solve the differential equation \eqref{2}, consider the following
eigenvalue equation: 
\begin{equation}
[\hg,\Gamma]=\lambda\Gamma.\label{eq:17}
\end{equation}
In this equation, $\hg$ can be treated as a superoperator acting
on $\Gamma$. To distinguish $\hg$ as a superoperator from that as
an operator, we denote the superoperator of $\hg$ as $\sh$, and
\eqref{eq:17} can be rewritten as 
\begin{equation}
\sh\Gamma=\lambda\Gamma.
\end{equation}

It is easy to verify that $\sh$ is an Hermitian superoperator (see
Appendix A). Therefore, $\sh$ has $d^{2}$ real eigenvalues, some
of which may be degenerate. Suppose the eigenvalues of $\sh$ are
$\lambda_{1},\cdots,\lambda_{d^{2}}$, and that $\lambda_{k}=0$ for
$k=1,\cdots,r$ and that $\lambda_{k}\neq0$ for $k=r+1,\cdots,d^{2}$,
and denote the corresponding orthonormal eigenvectors as $\Gamma_{1},\cdots,\Gamma_{d^{2}}$,
satisfying $\tr(\Gamma_{i}^{\dagger}\Gamma_{j})=\delta_{ij}$. Then,
$\partial_{g}\hg$ can be decomposed as 
\begin{equation}
\partial_{g}\hg=\sum_{k=1}^{d^{2}}c_{k}\Gamma_{k},
\end{equation}
where $c_{k}=\tr(\Gamma_{k}^{\dagger}\partial_{g}\hg)$. Since $Y(\mu)$
can also be decomposed in terms of $\Gamma_{1},\cdots,\Gamma_{d^{2}}$,
and $Y(0)=\partial_{g}\hg$, the solution of Eq. \eqref{2} is 
\begin{equation}
Y(\mu)=\sum_{k=1}^{d^{2}}\tr(\Gamma_{k}^{\dagger}\partial_{g}\hg)\e^{-\i\lambda_{k}\mu}\Gamma_{k}.
\end{equation}
Now, we can insert the above solution for $Y(\mu)$ into \eqref{eq:1-1},
and since the first $r$ eigenvalues of $\sh$ are zero, 
\begin{equation}
\begin{aligned}\h & =t\sum_{k=1}^{r}\tr(\Gamma_{k}^{\dagger}\partial_{g}\hg)\Gamma_{k}\\
 & -\i\sum_{k=r+1}^{d^{2}}\frac{1-\e^{-\i\lambda_{k}t}}{\lambda_{k}}\tr(\Gamma_{k}^{\dagger}\partial_{g}\hg)\Gamma_{k}.
\end{aligned}
\label{eq:18}
\end{equation}

Equation \eqref{eq:18} is the general solution for $\h$. When one
obtains the eigenvalues and eigenvectors of $\sh$ from \eqref{eq:17}
and plugs them into \eqref{eq:18}, $\h$ can then be derived.

If we know the eigenvalues and eigenstates of $\hg$ (as an ordinary
operator), the solution for $\h$ \eqref{eq:18} can be greatly simplified.
Suppose $\hg$ has $\n$ different eigenvalues, $E_{1},\cdots,E_{\n}$,
the degeneracy of $E_{k}$ is $d_{k}$, and the eigenstates corresponding
to $E_{k}$ are $|E_{k}^{(1)}\rangle,\cdots,|E_{k}^{(d_{k})}\rangle$.
The eigenvectors and eigenvalues of $\sh$ are 
\begin{equation}
\Gamma_{kl}^{(ij)}=|E_{k}^{(i)}\rangle\langle E_{l}^{(j)}|,\,\lambda_{kl}^{(ij)}=E_{k}-E_{l}.\label{eq:19}
\end{equation}

It is obvious that the degeneracy of the zero eigenvalue is $d_{1}^{2}+\cdots+d_{\n}^{2}$,
and the corresponding eigenvectors are $\Gamma_{kk}^{(ij)},\,i,j=1,\cdots,d_{k},\,k=1,\cdots,\n$.
The coefficients of these eigenvectors in $\h$ are 
\begin{equation}
\begin{aligned}\tr(\Gamma_{kk}^{(ij)\dagger}\partial_{g}\hg) & =\langle E_{k}^{(j)}|\partial_{g}\hg|E_{k}^{(i)}\rangle\\
 & =\partial_{g}E_{k}\delta_{ij}.
\end{aligned}
\label{eq:12}
\end{equation}
The eigenvectors with nonzero eigenvalues of $\sh$ are $\Gamma_{kl}^{(ij)},\,k\neq l$,
and their coefficients in $\h$ are 
\begin{equation}
\begin{aligned}\tr(\Gamma_{kl}^{(ij)\dagger}\partial_{g}\hg) & =\langle E_{l}^{(j)}|\partial_{g}\hg|E_{k}^{(i)}\rangle\\
 & =E_{k}\langle E_{l}^{(j)}|\partial_{g}E_{k}^{(i)}\rangle+E_{l}\langle\partial_{g}E_{l}^{(j)}|E_{k}^{(i)}\rangle\\
 & =(E_{k}-E_{l})\langle E_{l}^{(j)}|\partial_{g}E_{k}^{(i)}\rangle,
\end{aligned}
\label{eq:20}
\end{equation}
where we have used $\langle E_{l}^{(j)}|\partial_{g}E_{k}^{(i)}\rangle+\langle\partial_{g}E_{l}^{(j)}|E_{k}^{(i)}\rangle=\partial_{g}\langle E_{l}^{(j)}|E_{k}^{(i)}\rangle=0$.

By plugging \eqref{eq:19}-\eqref{eq:20} into \eqref{eq:18}, we
finally have 
\begin{equation}
\begin{aligned}\h & =t\sum_{k=1}^{\n}\frac{\partial E_{k}}{\partial g}P_{k}+2\sum_{k\neq l}\sum_{i=1}^{d_{k}}\sum_{j=1}^{d_{l}}\e^{\frac{-\i(E_{k}-E_{l})t}{2}}\\
 & \times\sin\frac{(E_{k}-E_{l})t}{2}\langle E_{l}^{(j)}|\partial_{g}E_{k}^{(i)}\rangle|E_{k}^{(i)}\rangle\langle E_{l}^{(j)}|,
\end{aligned}
\label{eq:1-2}
\end{equation}
where $P_{k}$ is the projection onto the eigensubspace corresponding
to $E_{k}$: $P_{k}=\sum_{i=1}^{d_{k}}|E_{k}^{(i)}\rangle\langle E_{k}^{(i)}|$.
We have used $1-\e^{-\i(E_{k}-E_{l})t}=2\i\exp\frac{-\i(E_{k}-E_{l})t}{2}\sin\frac{(E_{k}-E_{l})t}{2}$.

The form of $\h$ in \eqref{eq:1-2} implies that the quantum Fisher
information $\qf$ can be divided into two parts: one is due to the
dependence of the eigenvalues $E_{k}$ on $g$, and this part is linear
in the time $t$; the other is due to the dependence of the eigenstates
$|E_{k}^{(i)}\rangle$ on $g$, and that part oscillates with time.

When the dimension of the system is low, one may find the eigenvalues
and the eigenstates of the Hamiltonian explicitly, so Eq. \eqref{eq:1-2}
is a more direct and compact result for $\h$. However, if the dimension
of the system is very high, e.g., a condensed matter system, then
the eigenvalues and the eigenstates will be extremely difficult to
obtain, and the general result \eqref{eq:18} will be more helpful.
In this case, the eigenvalues and eigenstates of $\sh$ are still
unavailable, but one can get some knowledge of the quantum Fisher
information $\qf$ from the symmetry of the Hamiltonian.

For example, if $H$ is invariant under a unitary operation $U=\exp(-\i\Omega)$,
then $[\hg,\Omega]=0$, which implies that $\Omega$ is an eigenvector
of $\sh$ with eigenvalue zero. Thus one can calculate the coefficient
$\tr(\Omega\partial_{g}\hg)$ and check whether $\Omega$ belongs
to the support of $\partial_{g}\hg$. If it does, then the quantum
Fisher information $\qf$ will scale as $t^{2}$ when $t\gg1$. So
we can see that even lacking details about the eigenvalues and eigenvectors
of $\hg$, \eqref{eq:18} can give some information about the scaling
of $\qf$ through the symmetry of $\hg$.

\subsection{Upper bound on the quantum Fisher information $\protect\qf$}

From \eqref{eq:18} or \eqref{eq:1-2}, we can obtain an upper bound
on the quantum Fisher information $\qf$.

First, we note that 
\begin{equation}
\langle\Delta\h[2]\rangle_{\max}\leq\frac{1}{2}\tr(\h[\dagger]\h),\label{eq:21}
\end{equation}
(see Appendix B for a proof), so from \eqref{eq:13} and \eqref{eq:18},
we can derive 
\begin{equation}
\begin{aligned}\qf[,\max] & \leq2t^{2}\sum_{k=1}^{r}|\tr(\Gamma_{k}^{\dagger}\partial_{g}\hg)|^{2}\\
 & +8\sum_{k=r+1}^{d^{2}}\frac{|\tr(\Gamma_{k}^{\dagger}\partial_{g}\hg)|^{2}}{\lambda_{k}^{2}}\sin^{2}\frac{\lambda_{k}t}{2}.
\end{aligned}
\end{equation}
And when we know the eigenvalues and eigenstates of the Hamiltonian
$\hg$, the upper bound can be simplified to 
\begin{equation}
\begin{aligned}\qf[,\max] & \leq2t^{2}\sum_{k=1}^{n_{g}}d_{k}(\partial_{g}E_{k})^{2}\\
 & +8\sum_{k\neq l}\sum_{i=1}^{d_{k}}\sum_{j=1}^{d_{l}}|\sin\frac{1}{2}(E_{k}-E_{l})t|^{2}|\langle E_{l}^{(j)}|\partial_{g}E_{k}^{(i)}\rangle|^{2}.
\end{aligned}
\end{equation}

In particular, if the eigenvalues of $\hg$ are independent of $g$,
the upper bound of $\qf$ will not grow as $t^{2}$ when $t$ is large,
and the bound becomes 
\begin{equation}
\begin{aligned}\qf[,\max] & \leq8\sum_{k\neq l}\sum_{i=1}^{d_{k}}\sum_{j=1}^{d_{l}}|\sin\frac{1}{2}(E_{k}-E_{l})t|^{2}|\langle E_{l}^{(j)}|\partial_{g}E_{k}^{(i)}\rangle|^{2}\\
 & \leq8\sum_{k\neq l}\sum_{i=1}^{d_{k}}\sum_{j=1}^{d_{l}}|\langle E_{l}^{(j)}|\partial_{g}E_{k}^{(i)}\rangle|^{2}.
\end{aligned}
\end{equation}
In this case, the quantum Fisher information $\qf$ is always finite,
no matter how long the time $t$ is, in sharp contrast to the time
scaling of the Fisher information for estimating an overall multiplicative
factor of a Hamiltonian.

\section{Example: a spin-$\frac{1}{2}$ in a magnetic field}

In this section, we consider an example to illustrate the results
in the previous sections. We study the quantum Fisher information
in estimating a parameter of a magnetic field by measuring a spin-$\frac{1}{2}$
particle in the field.

Suppose the magnetic field is $B\overrightarrow{n_{\theta}}$, where
$B$ is the amplitude of the magnetic field and $\overrightarrow{n_{\theta}}=(\cos\theta,0,\sin\theta)$,
gives its direction. The parameter $\theta$ denotes the angle between
the direction of the magnetic field and the $z$ axis. Now we place
a spin-$\frac{1}{2}$ particle, e.g., an electron, in this magnetic
field and our task is to estimate the angle $\theta$ by measuring
this particle.

The interaction Hamiltonian between the the particle and the magnetic
field is 
\begin{equation}
H_{\theta}=B(\cos\theta\sigma_{x}+\sin\theta\sigma_{z}),
\end{equation}
where $\sigma_{x}$ and $\sigma_{z}$ are Pauli operators. We have
assumed $e=m=c=1$ in the above Hamiltonian for simplicity.

The eigenvalues of $H_{\theta}$ are $\pm B$, and the corresponding
eigenstates are 
\begin{equation}
|+B\rangle=\begin{pmatrix}\cos(\frac{\pi}{4}-\frac{\theta}{2})\\
\sin(\frac{\pi}{4}-\frac{\theta}{2})
\end{pmatrix},\,|-B\rangle=\begin{pmatrix}\sin(\frac{\pi}{4}-\frac{\theta}{2})\\
-\cos(\frac{\pi}{4}-\frac{\theta}{2})
\end{pmatrix}.
\end{equation}
According to \eqref{eq:1-2}, it can be obtained that the generator
$h$ of the local translation with respect to the parameter $\theta$
for an evolution of time $t$ is 
\begin{equation}
h=\begin{pmatrix}0 & \e^{-\i Bt}\sin Bt\\
\e^{\i Bt}\sin Bt & 0
\end{pmatrix},
\end{equation}
where the computational basis of $h$ is $|\pm B\rangle$. The eigenvalues
of $h$ are $\pm\sin Bt$, so the maximum quantum Fisher information
is 
\begin{equation}
F_{\max}^{(Q)}=4\sin^{2}Bt.\label{eq:25}
\end{equation}

We can also extend this result to a more general case. Suppose the
direction of the magnetic field $\overrightarrow{n_{\theta}}$ has
an arbitrary form with $\|\overrightarrow{n_{\theta}}\|=1$, then
the Hamiltonian of the interaction between the particle and the magnetic
field is 
\begin{equation}
H_{\theta}=B\overrightarrow{n_{\theta}}\cdot\overrightarrow{\sigma},
\end{equation}
where $\overrightarrow{\sigma}=(\sigma_{x},\sigma_{y},\sigma_{z})$
is the vector of the Pauli operators.

We can obtain $h$ from \eqref{eq:1-2}, 
\begin{equation}
h=\sin Bt\,(\cos Bt\overrightarrow{\partial_{\theta}n_{\theta}}-\sin Bt\overrightarrow{\partial_{\theta}n_{\theta}}\times\overrightarrow{n_{\theta}})\cdot\overrightarrow{\sigma}.
\end{equation}
Since $\overrightarrow{\partial_{\theta}n_{\theta}}\times\overrightarrow{n_{\theta}}$
is orthogonal to $\overrightarrow{\partial_{\theta}n_{\theta}}$,
and $\|\overrightarrow{\partial_{\theta}n_{\theta}}\times\overrightarrow{n_{\theta}}\|=\|\overrightarrow{\partial_{\theta}n_{\theta}}\|$,
the eigenvalues of $h$ are 
\begin{equation}
\pm\|\overrightarrow{\partial_{\theta}n_{\theta}}\|\sin Bt.
\end{equation}
Therefore, the maximum quantum Fisher information of estimating $\theta$
is 
\begin{equation}
F_{\max}^{(Q)}=4\|\overrightarrow{\partial_{\theta}n_{\theta}}\|^{2}\sin^{2}Bt.\label{eq:26}
\end{equation}

From \eqref{eq:25} and \eqref{eq:26}, we can see that the maximum
quantum Fisher information oscillates with the time $t$, and the
period of the oscillation is $\frac{\pi}{B}$. This implies that the
maximum quantum Fisher information is always bounded in this case,
and the upper bound is $4\|\overrightarrow{\partial_{\theta}n_{\theta}}\|^{2}$.
This is in sharp contrast to the case where the parameter to estimate
is an overall multiplicative factor of the Hamiltonian (compare to
the amplitude case below). In that case, the maximum quantum Fisher
information grows as $t^{2}$, and is unbounded as $t\rightarrow\infty$.

By way of comparison, if instead we want to estimate a parameter in
the amplitude $B_{g}$ of the magnetic field, where $g$ is the parameter
to estimate, and the direction of the magnetic field is fixed as $\overrightarrow{n}$,
then 
\begin{equation}
h=\partial_{g}B_{g}\overrightarrow{n}\cdot\overrightarrow{\sigma}.
\end{equation}
In this case, the maximum quantum Fisher information is 
\begin{equation}
F_{\max}^{(Q)}=4(\partial_{g}B_{g})^{2}t^{2},
\end{equation}
which recovers the time scaling $t^{2}$, which is known in quantum
metrology for phase estimation.

The maximum quantum Fisher information \eqref{eq:26} for estimating
$\theta$ has an intuitive physical picture. The derivative $\overrightarrow{\partial_{\theta}n_{\theta}}$
characterizes how fast the direction $\overrightarrow{n_{\theta}}$
changes with the parameter $\theta$. If $\overrightarrow{n_{\theta}}$
changes quickly with the parameter $\theta$, it will be more sensitive
to distinguish different $\theta$, and the precision of estimating
$\theta$ will be higher.

\section{Conclusion}

In summary, in this paper we studied quantum metrology for estimating
a general parameter of a Hamiltonian. We obtained the generator $\h$
of the infinitesimal parameter translation with respect to $g$, of
which the variance is the quantum Fisher information, and also a general
upper bound on the quantum Fisher information. The results show that
the optimal scaling of the quantum Fisher information with the number
of systems can always reach the Heisenberg limit, but the time scaling
can be different from that of estimating an overall multiplicative
factor. We considered estimating a parameter of a magnetic field by
measuring a spin-$\frac{1}{2}$ particle as an example to illustrate
the results, and compared estimating a parameter of the magnetic field
amplitude to estimating a parameter of the magnetic field direction.
When estimating a parameter of the magnetic field amplitude, the time
scaling of the quantum Fisher information is $t^{2}$, but when estimating
the parameter of the magnetic field direction, the quantum Fisher
information oscillates as a sine function of $t$. This example clearly
shows the difference between estimating an overall multiplicative
factor and estimating a general parameter, and gives a physical picture
illustrating the general results. 
\begin{acknowledgments}
The authors acknowledge the support from the ARO MURI under Grant
No. W911NF-11-1-0268 and NSF Grant No. CCF-0829870. 
\end{acknowledgments}

\section*{Appendix A: Proof of the hermicity of $\protect\sh$}

Suppose $\{\sigma_{1},\cdots,\sigma_{d^{2}}\}$ is an orthonormal
basis in the operator space; then the $(i,j)$th element of the superoperator
$\sh$ is 
\begin{equation}
(\sh)_{ij}=\tr(\sigma_{i}^{\dagger}[\hg,\sigma_{j}]),\label{eq:23}
\end{equation}
and 
\begin{equation}
(\sh)_{ij}^{\dagger},=\tr([\sigma_{i}^{\dagger},\hg]\sigma_{j}).\label{eq:24}
\end{equation}

If $\sh$ is Hermitian, it must satisfy $(\sh)_{ij}=(\sh)_{ij}^{\dagger}$.
We can check whether this is true directly from \eqref{eq:23} and
\eqref{eq:24}. Note that 
\begin{equation}
\begin{aligned}(\sh)_{ij}-(\sh)_{ij}^{\dagger} & =\tr(\sigma_{i}^{\dagger}[\hg,\sigma_{j}])+\tr([\hg,\sigma_{i}^{\dagger}]\sigma_{j})\\
 & =\tr([\hg,\sigma_{i}^{\dagger}\sigma_{j}])\\
 & =0,
\end{aligned}
\end{equation}
so this proves the Hermicity of $\sh$.

\section*{Appendix B: Proof of Eq. (\ref{eq:21}) }

First, we note that \cite{optimality 1} 
\begin{equation}
\langle\Delta\h[2]\rangle_{\max}=\frac{1}{4}(\lambda_{\max}-\lambda_{\min})^{2},
\end{equation}
where $\lambda_{\max}$ and $\lambda_{\min}$ are the maximum and
minimum eigenvalues of $h$, respectively.

On one hand, $|\lambda_{\max}-\lambda_{\min}|\leq|\lambda_{\max}|+|\lambda_{\min}|$,
so 
\begin{equation}
\begin{aligned}\langle\Delta\h[2]\rangle_{\max} & \leq\left(\frac{|\lambda_{\max}|+|\lambda_{\min}|}{2}\right)^{2}\\
 & \leq\frac{|\lambda_{\max}|^{2}+|\lambda_{\min}|^{2}}{2},
\end{aligned}
\label{eq:24-1}
\end{equation}
where the second inequality follows from the well-known \emph{power
mean inequality}: for any real positive numbers $x_{1},\cdots,x_{n}$
and nonzero $p,\,q$, 
\begin{equation}
\left(\frac{x_{1}^{q}+\cdots+x_{n}^{q}}{n}\right)^{\frac{1}{q}}\leq\left(\frac{x_{1}^{p}+\cdots+x_{n}^{p}}{n}\right)^{\frac{1}{p}},\,\text{if}\,p\geq q.
\end{equation}
If we take $q=1$ and $p=2$, it will produce \eqref{eq:24-1}.

On the other hand, 
\[
\tr(\h[\dagger]\h)=\sum_{k}|\lambda_{k}|^{2}\geq|\lambda_{\max}|^{2}+|\lambda_{\min}|^{2},
\]
where $\lambda_{k}$ runs over all eigenvalues of $\h$, so we have
\begin{equation}
\langle\Delta\h[2]\rangle_{\max}\leq\frac{1}{2}\tr(\h[\dagger]\h),
\end{equation}
which proves Eq. \eqref{eq:21}.

\end{document}